\documentclass[pdflatex,sn-mathphys-num]{sn-jnl}

\usepackage{graphicx}%
\usepackage{multirow}%
\usepackage{amsmath,amssymb,amsfonts}%
\usepackage{amsthm}%
\usepackage{mathrsfs}%
\usepackage[title]{appendix}%
\usepackage{xcolor}%
\usepackage{textcomp}%
\usepackage{manyfoot}%
\usepackage{booktabs}%
\usepackage{algorithm}%
\usepackage{algorithmicx}%
\usepackage{algpseudocode}%
\usepackage{listings}%
\usepackage{booktabs}
\usepackage{longtable}

\theoremstyle{thmstyleone}%
\theoremstyle{thmstyletwo}%

\theoremstyle{thmstylethree}%

\begin{document}

\title[Article Title]{The Lifetime of the Covid Memorial Wall: Modelling with Collections Demography, Social Media Data and Citizen Science}


\author*[1]{\fnm{Josep} \sur{Grau-Bové}}\email{josep.grau.bove@ucl.ac.uk}

\author[1]{\fnm{Mara} \sur{Cruz}}

\author[1]{\fnm{Pakhee} \sur{Kumar}}

\affil*[1]{\orgdiv{Institute for Sustainable Heritage}, \orgname{University College London}}


\abstract{The National Covid Memorial Wall in London, featuring over 240,000 hand-painted red hearts, faces significant conservation challenges due to the rapid fading of the paint. This study evaluates the transition to a better-quality paint and its implications for the wall's long-term preservation. The rapid fading of the initial materials required an unsustainable repainting rate, burdening volunteers. Lifetime simulations based on a collections demography framework suggest that repainting efforts must continue at a rate of some hundreds of hearts per week to maintain a stable percentage of hearts in good condition. This finding highlights the need for a sustainable management strategy that includes regular maintenance or further reduction of the fading rate. 

Methodologically, this study demonstrates the feasibility of using a collections demography approach, supported by citizen science and social media data, to inform heritage management decisions. An agent-based simulation is used to propagate the multiple uncertainties measured. The methodology provides a robust basis for modeling and decision-making, even in a case like this, where reliance on publicly available images and volunteer-collected data introduces variability. Future studies could improve data within a citizen science framework by inviting public submissions, using on-site calibration charts, and increasing volunteer involvement for longitudinal data collection. This research illustrates the flexibility of the collections demography framework, firstly by showing its applicability to an outdoor monument, which is very different from the published case studies, and secondly by demonstrating how it can work even with low-quality data.}

\keywords{collections demography, memorial, fading, lifetimes}

\maketitle

\section{Introduction}\label{sec1}

This paper argues that collections demography can be effectively applied to outdoor monuments using low-cost data from citizen science and social media. The National Covid Memorial Wall, located opposite the Houses of Parliament in London, has two main conservation concerns: preventing its hearts from fading and finding a long-term sustainable management strategy. These two concerns are related, because the expense of maintenance depends on the rate of fading. With more than 240,000 red hearts painted on it, the memorial stands as a visual tribute to the lives lost during the pandemic. Each heart in the mural symbolises a human loss directly attributed to COVID-19 in the United Kingdom \cite{NationalCovidMemorialWall2024}.  Currently, due to weathering, the hearts are losing their red colour, which is noticeable along the entire memorial wall. The process of paint fading entails a weakening of colour saturation due to light exposure or chemical reactions, such as those caused by sunlight or rain \cite{Weyer2015}.  Understanding the lifetime of the painted hearts is of special interest to the volunteers currently managing and fighting for the preservation of the wall. Each week, a group of volunteers from the Covid-19 Bereaved Families for Justice meet to repaint the hearts that have faded due to environmental factors \cite{Booth2021}.  This work is significant, as it requires substantial efforts given the large number of hearts and the size of the wall. Therefore, in this paper, we follow a heritage science modelling approach to examine the ageing process and how different management strategies results in future rates of fading of the Covid Memorial Wall.
 
Heritage collections can be defined by a lifetime \cite{Strlic2013}. The concept of ‘collections demography’, which was introduced by Strlič et al. \cite{Strlic2013Demography},  involves the systematic study of the ageing of collections. Within this framework, collections are treated as populations of objects, wherein the focus is on assessing how different pre-existing conditions and preventive measures impact their lifetimes\cite{DuranCasablancas2021}.  This approach demonstrates significant potential as a tool to support decision-making in heritage practice, as evidenced by the work of Duran-Casablancas et al. \cite{DuranCasablancas2024},  which explores archival collections using agent-based modelling. In addition, recent policy developments, specially in relation to Cultural Capital in the UK, are emphasizing the benefits of extending damage modelling to different heritage assets \cite{dcms2024embedding}. However, the application of collections demography to new collection types is limited by the resources, data and time that are required. The basic steps of this framework (which are reviewed in section \ref{sec:colldem}, and are modelling change, understanding values, surveying collections, and predicting lifetimes) have mostly been conducted within the context of funded academic research projects. Collections demography, as described in the existing literature, requires extensive data, both on the nature of chemical change and on the condition of the collection under study. Therefore, the complexity of the collections demography framework is a significant constraint on a method that has a great potential to inform decision-making in the heritage field.
 
This paper challenges this limitation by presenting a comprehensive collections demography workflow, conducted at no cost by using mostly publicly available data and citizen science (CS). In simplified terms, CS involves the participation of nonexperts in scientific research \cite{Haklay2013}.  Participation in CS projects enables individuals to contribute to science through their intellectual efforts and knowledge, or by providing resources \cite{Serrano2014}.  A related concept is Crowdsourcing, an umbrella term for a variety of approaches in which a large group of people perform small tasks in order to achieve a collective goal. Numerous studies have demonstrated the adaptability of CS and crow sourcing to different scales and subjects of study many fields, including heritage, for example by monitoring threatened heritage sites \cite{Brigham2022}, counting birds to understand population dynamics during a pandemic \cite{Coldren2022}, classifying galaxies \cite{Raddick2019}  or measuring light pollution at night \cite{Nugent2017}, to cite some examples. 

The prominence of the National Covid Memorial Wall, and the wide availability of photographs of the hearts in social media, present an ideal scenario for developing a low-cost collections demography workflow than can inform its management. In addition to being a highly visible public monument, the physical properties of the wall are also well-suited to collections demography. It hosts a “collection” of hundreds of thousands of hearts, with similar properties, and which age in a similar way. The ageing process, fading, can easily be measured with a clear metric, colour. The time-scale of fading is such that short, multiple-month experiments can capture change. It also benefits from the involvement of a highly motivated group of volunteers who are eager to collaborate with scientists and actively explore various preservation strategies during the research period. Therefore, this paper aims to:

 \begin{itemize}
    \item present a complete low-cost collections demography workflow that relies on CS and social media;
    \item study, through the case of the National Covid Memorial Wall, the benefits and limitations of this approach; and
    \item use the resulting model to examine possible management scenarios for the National Covid Memorial Wall.
 \end{itemize}
	
After providing an overview of the memorial and the fading process of its painted hearts, the paper details the methodology behind the fading measurements, including data collection, image calibration, and citizen science experiments. The research then explores the acceptability of fading through surveys and introduces a series of agent-based models to evaluate various management strategies for the memorial. The paper concludes by analysing the effectiveness of different repainting strategies, discussing their potential costs and benefits, and offering recommendations for sustainable long-term maintenance. This structure aims to demonstrate the flexibility of a collections demography approach, and how it can be applied to unique heritage objects, specially if they share some of the features of National Covid Memorial Wall.

    \begin{figure}[ht]
\centering
\includegraphics[width=0.9\textwidth]{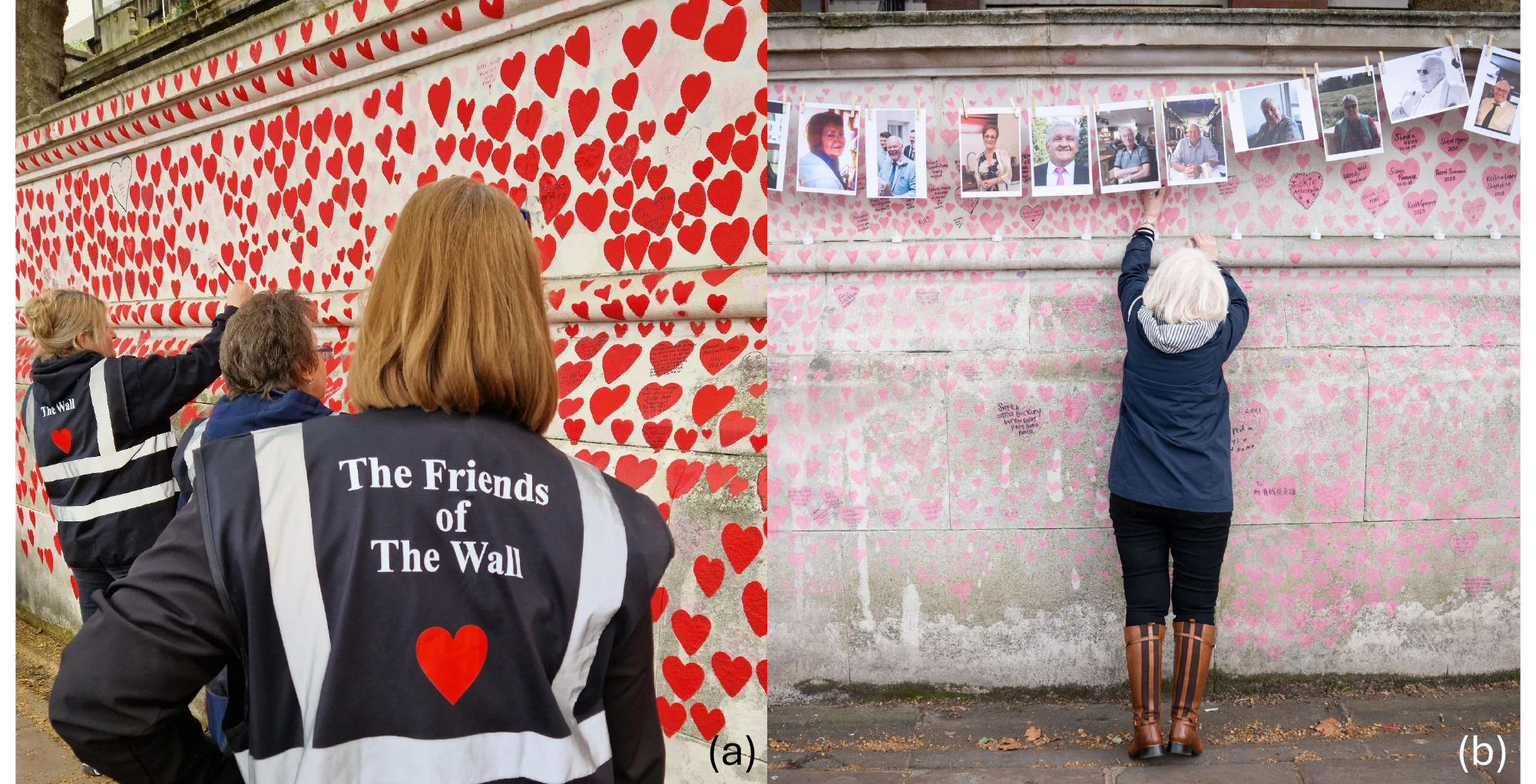}
\caption{(a) A bereaved volunteer working on repainting and re-inscribing the names of the deceased on the faded hearts, which were originally painted with red Posca markers.  (b). The Friends of the Wall repaint only the hearts with legible inscriptions using premium red Valspar paint (Source: National Covid Memorial Wall UK Instagram, 21 May 2023)}\label{fig1}
\end{figure}

    \section{Understanding the National Covid Memorial Wall}
   
The National Covid Memorial Wall is a public mural featuring over 240,000 painted red hearts (July 2024), which are gradually fading. Each heart is dedicated to the memory of loved ones who passed away from COVID-19 \cite{NationalCovidMemorialWall2024}.  The memorial stretches 500 metres along the South Bank of the River Thames, from Westminster Bridge to Lambeth Bridge, directly opposite the Houses of Parliament. Established in 2021, it was created as a protest piece by two groups: COVID-19 Bereaved Families for Justice and Led By Donkeys \cite{Lynskey2021}.  The wall is owned by Guy’s and St Thomas’ NHS Foundation Trust, and the walkway is maintained by Lambeth Borough Council.

Each week, the Friends of the Wall,  group of bereaved volunteers, manually add or repaint hearts on the memorial. Much of the information in this section comes directly from members of this group. They are among the key stakeholders in the memorial's preservation, along with other bereaved individuals who maintain hearts with which they have a personal connection, and institutional actors such as the local council and nearby hospital. In addition to their caretaking role, members of the Friends of the Wall are active campaigners: they are invested in the Covid hearings, engage with policymakers, and regularly host events at the wall, that have been atteded by Members of Parliament.
 
The memorial holds a multiplicity of values for different publics. For bereaved families and friends, it represents personal and emotional connection to lost loved ones. For the broader public, it is a symbol of collective grief and solidarity. Campaigners value the wall as a site of political expression and public accountability, particularly in relation to government responses to the crisis. It also has political value: in 2021, a letter to the Prime Minister Boris Johnson signed by 227 Members of Parliament, Peers and Mayors, asked for the wall to be converted into a permanent memorial \cite{Khan2021CovidMemorialWall}. It has been mentioned 47 seven times so far in Parliament according to the Hansard, the latest in March 2025, when expressing the need to pay tribute to the bereaved, Baroness Twycross said "We have a visual reminder of that opposite Parliament in the form of the Covid memorial wall" \cite{UKParliament2025}.
 
While much should be written about its value, this paper focuses mostly on its materiality. The memorial is crafted from three materials: Portland limestone for the masonry construction, red Posca paint markers or red Valspar Premium Masonry paint for the hearts and black Sharpie pens for inscribing the names of the deceased within the hearts. In the process of repainting, stencils are not used, rather, the volunteers prefer to apply paint in a more controlled way and respect the variability of the shapes of the hearts. The volunteers also repaint old hearts, rewrite the names inscribed within the hearts when readable, and remove graffiti. White Valspar Premium Masonry paint is also used to alter the shapes of hearts that are seen to exceed the agreed size.

Initially, red Posca paint markers were used to paint the hearts in 2021. However, these markers faded rapidly, leading to a second phase of painting in 2022 with red Valspar Premium Masonry paint instead. Figure \ref{fig1}a illustrates the extent of fading with the red Posca markers, including the point at which the name inscriptions became illegible. At the time of submission of this paper, the hearts are being repainted with Valspar Premium Masonry paint, with only those hearts with illegible inscriptions remaining unchanged, as shown in Figure \ref{fig1}b. 

The most pressing conservation concern for the National Covid Memorial Wall is, arguably, preventing the hearts from fading. Being outdoors exposes the memorial to various environmental factors, including rainfall, humidity, and sunlight, which contribute to its gradual degradation. Additionally, its extensive nature (measuring 500 metres and comprising 240,000 hearts) requires significant weekly efforts from the Friends of the Wall in repainting the entire wall. Therefore, securing government protection as a permanent monument has become a key maintenance objective \cite{Vernon2024}.  The cost of repainting is a key consideration in finding a sustainable management solution. This cost depends on different factors, such as the fading rate of the paint and the number of hearts repainted. It also depends in a smaller measure on the strategy adopted to repaint the wall, as this paper will illustrate.

\section{Methodology: the elements of Collections Demography}
\label{sec:colldem}

\begin{figure}[ht]
\centering
\includegraphics[width=0.8\textwidth]{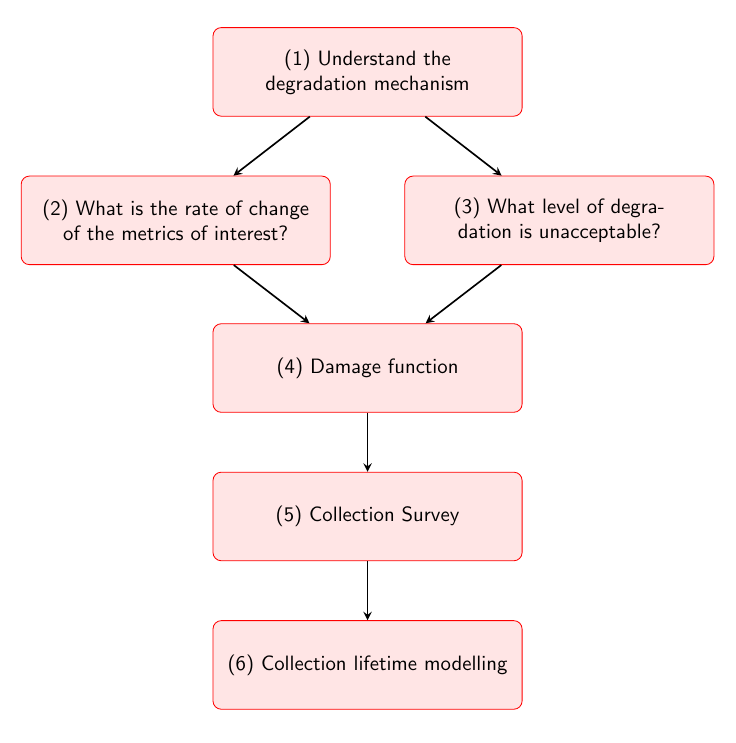}
\caption{The main steps of a collections demography framework. }\label{fig2}
\end{figure}

The collections demography workflow is illustrated in Figure \ref{fig2}.  All the steps conducted by Strlic and others \cite{Strlic2013Demography, DuranCasablancas2021} have been conducted in this research, but inserting social media data and citizen science when possible.  In the description that follows, we describe the conventional collections demography workflow and how it has been adapted in this work. We also comment, where applicable, which parts of the workflow could benefit from larger-scale crowdsourcing approaches in future work.  Table \ref{tab1} summarises how the method has been adapted. 

\begin{longtable}{@{}p{2.5cm}p{3.8cm}p{4.2cm}p{4.5cm}@{}}

\toprule
\textbf{Step} & \textbf{Conventional Method} & \textbf{This Study’s Adaptation} & \textbf{Future Potential (Citizen Science \& Crowdsourcing)} \\
\midrule
1. Understand degradation mechanisms & Laboratory analysis of material degradation (e.g., light ageing, chemical reactions) & Field observations of paint fading & Citizen monitoring of environmental factors (e.g., apps to measure UV, humidity) \\
\addlinespace
2. Determine rate of change & Controlled ageing tests or long-term expert monitoring & Colour change from publicly shared social media images, calibrated with on-site reference & Use of calibration charts on-site; improved quality via guided public image submissions \\
\addlinespace
3. Define unacceptable degradation & Expert visual judgement or consensus & Online survey measuring public perception of fading (\(\Delta E\) threshold) & Broader public input via crowd sourcing or on-site mobile prompts; multiple user group perspectives \\
\addlinespace
4. Build damage function & Empirical models based on physical measurements & Linear model of \(\Delta E\) increase from image analysis & Explore nonlinear or environmentally adjusted models; stakeholder co-development workshops \\
\addlinespace
5. Collection condition survey & Expert inspections using visual scales and imaging & Social media image analysis and citizen-collected measurements & Regular crowdsourced submissions; use of simplified mobile apps \\
\addlinespace
6. Lifetime modelling & Agent-based or deterministic models using expert-collected data & Agent-based simulation with public data and uncertainty propagation &  \\
\bottomrule
\caption{Adaptation of the collections demography framework using social media and citizen science.}
\label{tab1}
\end{longtable}

The Collections Demography process begins by understanding degradation mechanisms. This step involves identifying factors that cause deterioration and understanding their interaction with the materials of interest (Step 1). The next step is determining the rate of change of relevant metrics (Step 2). For example, in the case of paper collections, this would be the degree of polymerisation, or in the case of PVC, colour \cite{Rijavec2023}. In the case of the hearts on the wall, colour is also natural choice. The typical methods to determine the rate of change include accelerated ageing in a laboratory setting or direct monitoring, which is the approach taken in this case. Citizen science can contribute to Step 2 by providing observations of the parameters of interest. In this paper, the observations are obtained with a combination of citizen science and social media data. 

Step 3, "What level of degradation is unacceptable?", focuses on setting thresholds for acceptable deterioration, often based on expert judgment or the perception of relevant audiences. In this paper, Step 3 is performed with an online survey. Hypothetically, it could also benefit from crowdsourcing, meaning that the respondents of the survey do not necessarily have to be “subjects” to be studied, but micro-collaborators in the research. This methodology would allow for a wide range of perspectives to be heard to define these limits based on how fading is perceived in different contexts.

The data from Steps 2 and 3 are integrated into Step 4 to create a damage function, which predicts lifetimes based on rates of change and a threshold of unacceptable change \cite{Strlic2013}. Step 5, "Collection Survey", provides the condition of collections as an input to the following steps. It involves obtaining data on the current state of the collection. In a collection demography framework, the primary objective of this survey is understanding the average condition as well as the distribution of the input parameters. In other words, understanding the spread of condition of the collection. Surveys often use visual inspection, imaging techniques, and condition reporting scales. In this step, crowdsourcing can hypothetically be used to engage non-experts to participate in condition assessments. In this case, a combination of field measurements by citizen scientists and pictures from social media are enough to determine the distribution of fading at time zero. Finally, in Step 6, "Collection lifetime modelling," mathematical or computational models predict the future degradation of collections based on the damage function and collected survey data. 

\section{Measuring the fading rate of the hearts}

\subsection{Data collection} 

Data collection to measure the fading rate of the hearts on the National Covid Memorial Wall was conducted using various online sources, collecting freely available images. The selection criteria was to use images with accurate dating and which covered a period during which the hearts exhibited visible changes in colour to support subsequent analysis. A total of 426 images were collected from social media platforms, with upload dates between 30 March 2021 and 24 January 2023.  The platforms included Instagram (97 images), Twitter (39 images), Flickr (29 images), Red (26 images), Alamy (122 images), AP (24 images), Getty Images (39 images), Google Maps (14 images), and various other media sources (36 images). Given that the images were not reproduced, but simply analysed and colour extracted, the license of the images was not considered as a selection criteria. This number of images was comparable to the number of submissions obtained in crowd-sourcing projects by placing a sign on-site requesting visitors for photographs \cite{Brigham2022}, which is an approach that would probably produce good results in this case. 

Photographs with insufficient resolution (less than 752x501 pixels), those that were overexposed, filtered or imprecisely dated (for example, with a month rather than a day) were also excluded from the dataset. The accuracy of the date was checked by comparing the publication data of the social media post with available metadata, which was used when available. Given the large number of hearts, only images that featured the black board on the wall with the inscription 'National Covid Memorial Wall' were selected. The black board fulfilled two key functions: facilitating the identification of hearts within the same image and providing a calibration reference, as its colour was known to the researchers.

\subsection{Calibration of social media images}
    
Image calibration techniques were employed to precisely analyse the colours of the hearts at the National Covid Memorial Wall. First, the colour of the black board at the memorial site was measured using a colorimeter (an X-Rite 518 Colour Reflectance Spectrophotometer). This measurement was then used as a reference to adjust the colour balance of the images, ensuring alignment with the on-site readings of the black board. This calibration process is a simplified version of what has been conducted in the past with colour charts \cite{Brigham2018}, which would be also an option in this site if the research was repeated with appropriately designed signage. 

Subsequently, to validate the accuracy of the colour measurements obtained from the images, hearts with different levels of fading were selected on a specific day. The resulting colours were measured using a colorimeter, obtaining colour coordinates within the CIE L* a* b* system (CIELAB Colour Space), which provides a three-dimensional representation of the colour. These measurements were then correlated with photographs taken on the same day to assess the reliability of the photographs as a source of colour data. Seven smartphones (Samsung S21, iPhone 11, iPhone 12, iPhone 12 Pro Max, iPhone 13, iPhone 13 Pro, iPhone 14 Pro) captured images of the hearts and the black board, from varying distances and angles. Reference measurements were taken on site with the colorimeter, which provided average LAB values from three points on each heart. The smartphone images were processed using Python (specifically the cv2.color\_BGR2LAB function) to convert them from BGR to LAB format.

\begin{figure}[ht]
\centering
\includegraphics[width=0.6\textwidth]{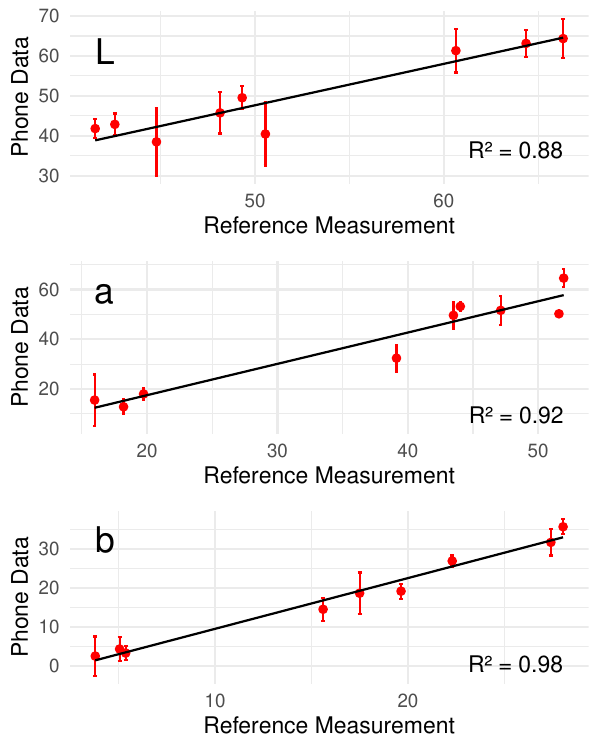}
\caption{The linear regression illustrates the high correlation between the L* a* b* values from the phone data and the reference measurements.}\label{fig3}
\end{figure}

Next, the L, a, and b coordinates from the images of a specific heart were compared with the reference measurements. The results, as shown in Figure \ref{fig3}, demonstrated a satisfactory level of correlation between the image data and the actual colour values of the hearts. Research has shown that better correlations can be obtained with more sophisticated adjustment of the images if color charts are present \cite{BarberoAlvarez2021}, which is a step that can be implemented in a site like this in the future. The quality of this correlation was deemed sufficient (in particular in the a dimension, which includes the colour red), to continue the research by correcting the colour of the collected photographs.

\subsection{Citizen Science experiments and current condition}

An additional low-cost colorimeter (Datacolor ColorReader) was used by citizen scientists, alongside the comparison between the research-grade colorimeter and social media photographs described above. This task did not produce many statistically useful results, but it is included in the methodology because it offers an opportunity to discuss the role of citizen science in this context. The low-cost colorimeter was used by volunteers of the Covid-19 Bereaved Families for Justice during their wall repainting sessions. A short in-situ training session was organised in November 2022. The measurements obtained were useful to determine the distribution of colour of the hearts at the start of the fading process. This experiment determined that the fresh paint has values of L = 49.3 b = 46.3 and c = 20.5 (the corresponding RGB is 194, 80 and 85). It was also found that, at the time of the measurement, the colour of the hearts is approximately distributed around $\Delta E$ = 5 of the mean.  After these measurements, the volunteers were encouraged to decide how and when to carry out the measurements during the following months. 

Apart from the useful on-site measurements taken in a single day, the main outcome of citizens science in this case was enabling a two-way collaboration, where the citizen scientists led the decision-making on which hearts to select. This led to useful conversations about which sections of the wall were fading faster, how often monitoring should happen, and how the personal value of some hearts influenced decisions on what to monitor. Although the data was not used in the model, the process enriched the project. To make the data statistically useful, more research time would need to be invested in strengthening the data collection protocols.
	
\subsection{Obtaining time-series of fading}

\begin{figure}[ht]
\centering
\includegraphics[width=0.9\textwidth]{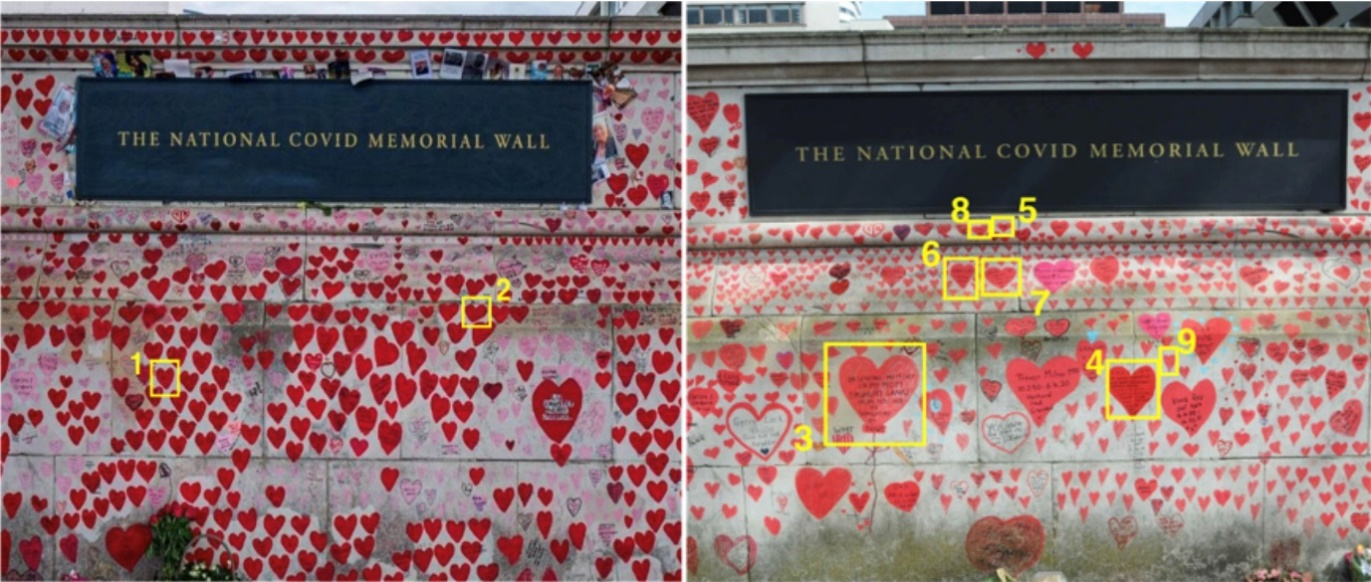}
\caption{The nine selected hearts that were tracked in social media images. Subsequently, the L* a* b* values (three-dimensional colour representation) and $\Delta E$ (quantitative measure of the difference between two colours) were calculated for each period corresponding to the selected hearts. }\label{fig4}
\end{figure}

By measuring the fading of the hearts from photographs of the Covid Memorial Wall, we generated time-series data showing the evolution of their colour. To achieve this, nine representative hearts located near the black board were selected and tracked through images on social media, as illustrated in Figure \ref{fig4}. The $\Delta E$ was calculated as the difference between the LAB values extracted from each photograph and the starting value measured at the wall on fresh paint.

 \begin{figure}[ht]
\centering
\includegraphics[width=0.9\textwidth]{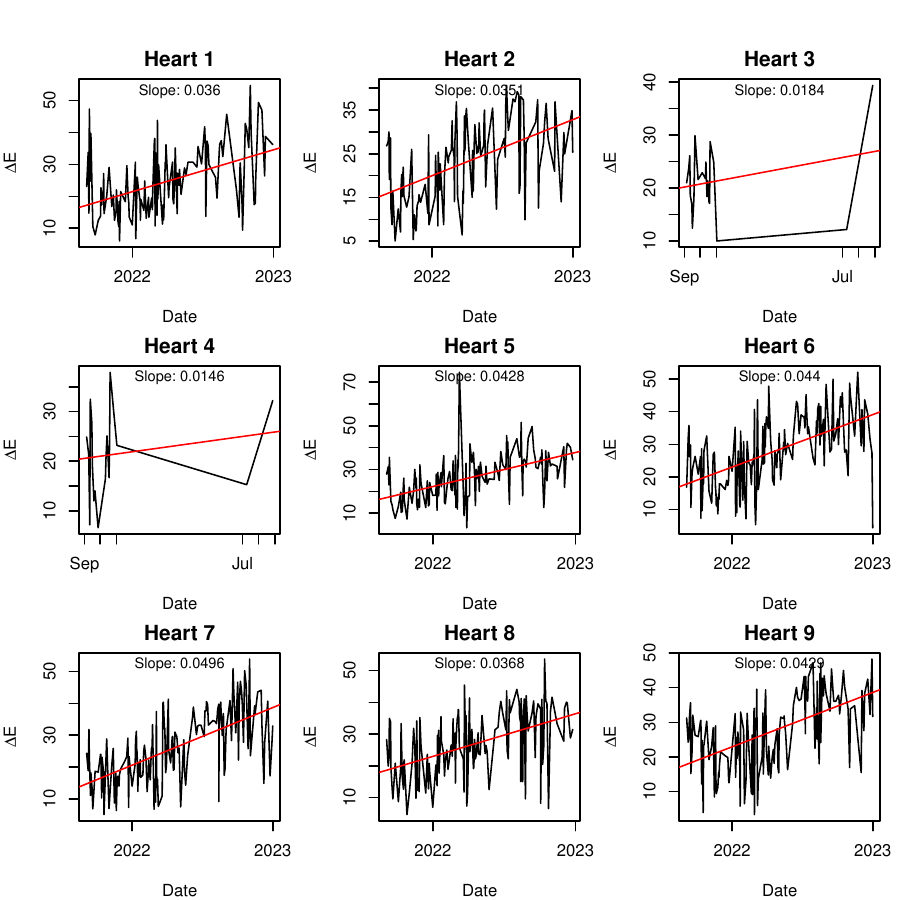}
\caption{Trend analysis, showing how the colour of the hearts has changed since their initial appearance in a photograph. Two hearts (3 and 4) didn't appear in enough good quality photographs to enable a regression. The time frame for Hearts 3 and 4 is September 2021 to July 2022}\label{fig5}
\end{figure}
 
\begin{figure}[ht]
\centering
\includegraphics[width=0.9\textwidth]{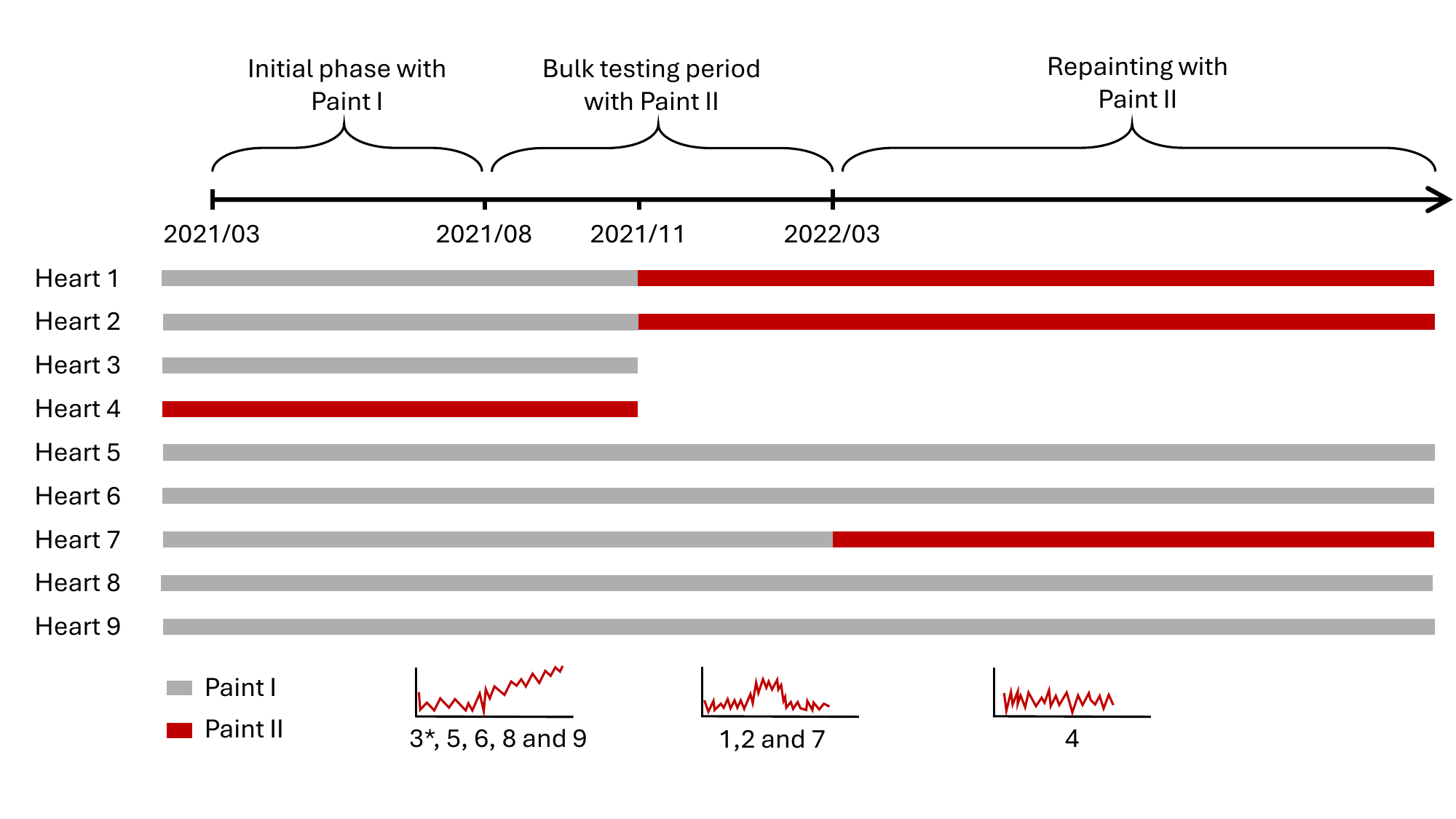}
\caption{Analysis of the nine representative hearts, illustrating the extent of change since their initial appearance in a photograph. The regression was conducted on the times marked in grey. The spark lines at the bottom summarise the fading trends of different hearts. The asterisk (*) indicates two hearts without sufficient data to conduct a regression.}\label{fig6}
\end{figure}

Among the nine tracked hearts, seven (Hearts 1, 2, 5, 6, 7, 8 and 9) display a reasonably consistent trend. Despite considerable measurement uncertainty, an upward trend in $\Delta E$ values is observed, indicating a gradual increase in discolouration from the original colour of the hearts to their current state (Figure \ref{fig6}). However, two hearts (Hearts 3 and 4) do not follow this trend due to insufficient data, as the available photographs featuring them before they were repainted only covers the period from September 2021 to July 2022. There are multiple error sources: the diversity of cameras used, potential misdating of the photographs, the diversity of times and lighting conditions of the photographs and the natural variability in colour and fading rate.

A linear regression was performed between $\Delta E$ and the number of days, for the periods where fading was detected, indicated in grey in Figure \ref{fig6}. An interesting observation is that, although the data exhibit some level of noise, the linear regression analysis reveals relatively consistent slopes, ranging from 0.0351 to 0.0496 $\Delta E$ units per day. This analysis provides an average rate of fading. Averaging the rate for the seven hearts available for the analysis, it is possible to calculate the error in the estimation of the rate, resulting in standard deviation of 0.0052 $\Delta E$ per day. Therefore, it can be determined that the hearts discolour at an approximate rate of 0.041 $\Delta E$ units per day, with an error of ±12\%. This finding is useful in the subsequent modelling of the fading and its uncertainty.

\subsection{Some observations on the fading data}

The analysis of the tracked hearts at the National Covid Memorial Wall provided valuable data on the fading of the paints used. Figure \ref{fig7} presents the nine selected hearts, showing the initial application of the paint and the bulk tests conducted with two types of paint: Paint I (Posca markers, shown in grey) and Paint II (Valspar Premium Masonry paint, shown in red).

Based on the data obtained, it is possible to identify which hearts faded rapidly and which did not, allowing for a hypothesis regarding the type of paint used in each case. In most of the hearts, the impact of the paint type is evident. Hearts painted in 2021 with Paint I faded quickly. By contrast, those repainted more recently showed a slower fading rate, consistent with the introduction of Paint II in 2022. This trend led us  to focus the regression analysis on the hearts and time periods that correspond to Paint I, as these showed significant changes over time. Paint II did not show sufficient variation during the selected period to calculate an accurate rate of fading. This may be attributed to the fact that, during the analysis period, the changes were too small to be detected with precision given the measurement noise. It is worth noting that this method is particularly effective for measuring fading in objects that undergo visible discolouration in a matter of months up to a year, but less suited for those showing slower changes. 

Figure \ref{fig6} displays the nine selected hearts at the National Covid Memorial Wall, highlighting the initial paint application of the two paint types. The timeline indicates the periods when the volunteers were experimenting with the new paint with groups of hearts, before deciding to use it permanently for repaints.

\section{Understanding the acceptability of change}

An essential element in the study of collection demography is the ability to assess both the rate at which changes occur and the point at which such changes become unacceptable. This is a necessary step for defining the lifetime of heritage assets. When an object undergoes changes that exceed acceptable limits, it is considered to have reached the end of its useful life and, therefore, requires intervention. In this context, a key step was determining the level of colour change at which stakeholders or visitors to the National Covid Memorial Wall perceive that the hearts need to be repainted. One of the volunteers, in conversation, was very clear with their criteria: the wall should always look red to those looking through the windows of the Houses of Parliament. The following steps outline a methodology to define the threshold in more numerical terms. 

For this analysis, five hearts were selected from a previously identified sample. Photographs of these hearts were selected corresponding to different time periods from the dataset of social media images, enabling the tracking of their colour evolution over time. With this information, an online survey was conducted, which attracted 150 participants. No demographic data was collected. In the survey, images of the hearts were shown, and beneath each image, the following question was asked: ‘I believe this heart is in need of repainting’. Respondents answered using a 5-point scale, where 1 indicated ‘Strongly disagree’, 2 ‘Somewhat disagree’, 3 ‘Neither agree nor disagree’, 4 ‘Agree’, and 5 ‘Strongly agree’.

 \begin{figure}[ht]
\centering
\includegraphics[width=0.9\textwidth]{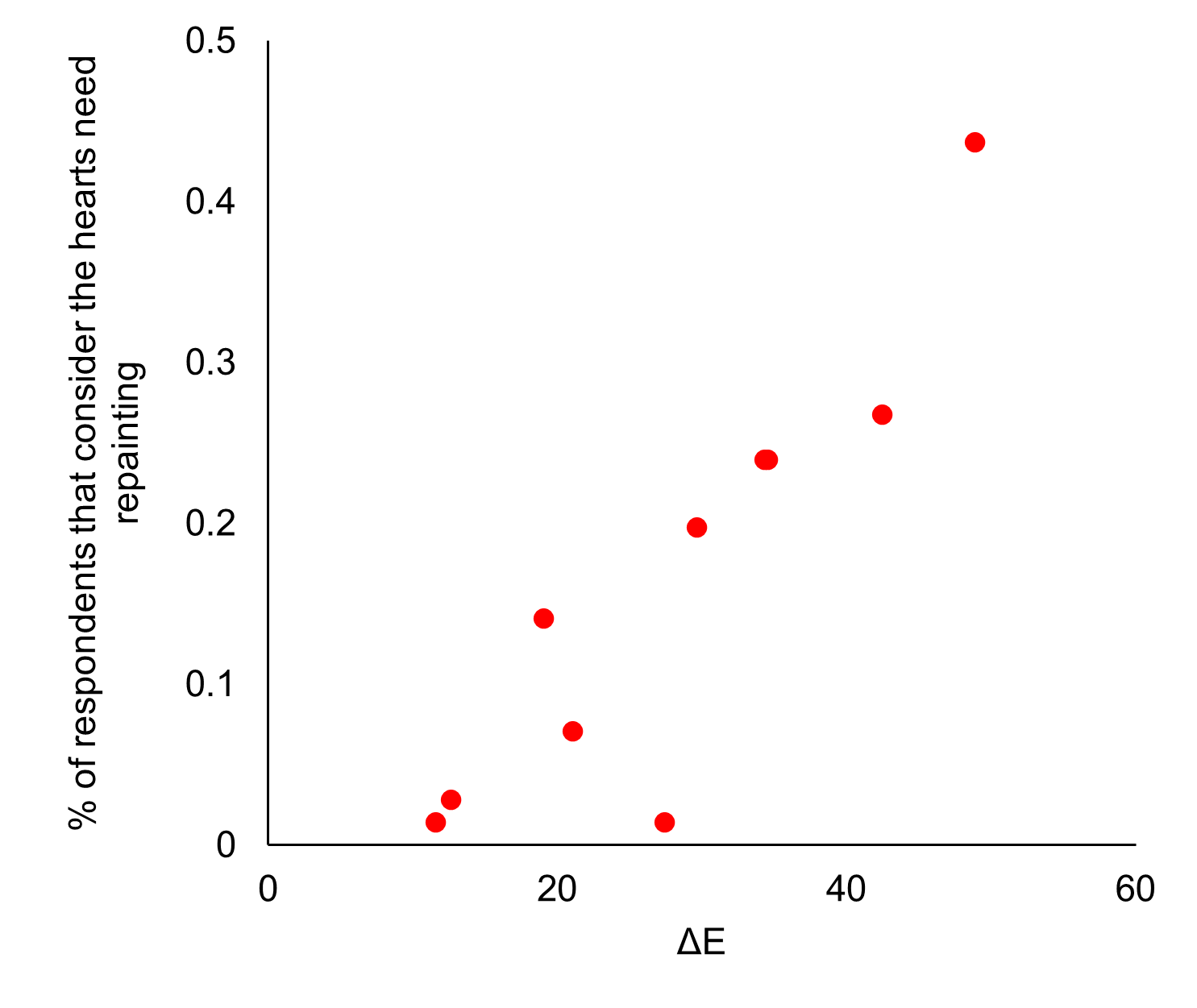}
\caption{ Relationship between heart colour and the percentage of visitors who agreed that repainting is needed.  }\label{fig7}
\end{figure}

Figure \ref{fig7} was generated based on these results, illustrating the relationship between the colour of the hearts and the percentage of visitors who either ‘agreed’ or ‘strongly agreed’ with the need to repaint. The analysis revealed a clear correlation: as more colour is lost, the number of visitors who consider repainting necessary increases.

It is important to note that this relationship is gradual, with no clear point of unanimous agreement on when repainting should occur. In other contexts, researchers have reported “tipping points”, for example where most observers agree a soiled surface looks too darkened \cite{Bellan2000}. As the hearts fade, more visitors support repainting, but it is not possible to identify a single threshold that would satisfy all. To achieve such consensus, it would be necessary to maintain an extremely low level of fading, below a $\Delta E$ of 10, as shown in Figure \ref{fig7}. This is particularly relevant given that the minimum colour difference perceptible to the human eye under ideal conditions, when two colours are placed side by side, is approximately $\Delta E$ 2-3.

It can be hypothesised that several factors contribute to the perceptual resistance to the fading of the hearts. Among these factors are their outdoor exposure, the variability in heart colours, the contrast between the examined hearts and the others that have been seen by participants, fluctuations in lighting conditions and the strong contrast provided by the pale Portland stone on which they are painted. These elements, combined, seem to minimise the perception of fading. Indeed, up to a $\Delta E$ of 20, very few observers wish the hearts to be repainted.

This information is useful for guiding decision making regarding the optimal time to repaint the hearts. If the goal is to maintain an optimal state, with vibrant colours that satisfy most visitors, it is recommended to keep the $\Delta E$ close to 20. However, it may be considered acceptable for up to 20\% of observers to perceive some level of fading, which would allow postponing repainting until a $\Delta E$ of 30 is reached, without significantly compromising the overall perception of the hearts’ condition. Each of these decisions will have a different impact on the number of hearts to be repainted, which is examined in the following sections. 

\section{Agent-based modelling of the collection}

Agent-based simulations are particularly well-suited for modelling cultural collections. They enable the representation of the degradation processes at the level of individual collection items, while also capturing the aggregate behaviour of  the system. This approach is useful for handling uncertainty, as it allows for multiple simulations using probabilistic sampling to explore the distribution of potential outcomes. For example, in the case of the Memorial Wall, agents can be initialized using the distribution of initial colours recorded by citizen scientists, and assigned degradation rates obtained from the social media pictures.

\subsection{Model Setup}

All the information collected is fed into an agent-based simulation model, which predicts how each heart will evolve in the future, including fading and eventual repainting. The model uses 1.000 agents, each of which represent a few hearts whose visible fading is described by a Delta E value. This value increases over time due to environmental exposure, following a linear degradation function: 

\begin{equation}
\Delta E = kt+C    
\end{equation}

where $k$ is the degradation rate, $t$ the time in days, and $C$ the initial value. Therefore, the equation simply models a gradual fading over time. The parameter $k$ is sampled from a normal distribution based on estimated degradation rates, and the model includes an error propagation term to account for uncertainty. 
A key feature of the model is its inclusion of perceptual and decision-making thresholds. A heart is perceived to need repainting once its Delta E exceeds a perception threshold (set at 10). This abstraction captures the public or stakeholder's sense of visual degradation. Three repainting strategies are simulated, as described below In all cases, repainting resets the Delta E of selected agents to zero.
Simulation results are tracked over 6,000 time steps in the longest simulation (approx. 16 years), and the percentage of agents perceived to require repainting is recorded over time. A control scenario with no repainting is also included to highlight the accumulation of visual degradation. Uncertainty is addressed by simulating upper and lower bounds using variation in the degradation rate k. 
The model allows users to control several parameters to explore different maintenance and degradation scenarios. These include the number of agents, representing the sample size of painted elements; the perceptual threshold, which defines the $\Delta E$ value above which a heart is perceived as needing repainting; and the daily repainting capacity, expressed as a fraction of the total population, to simulate different resource constraints.

\subsection{Management scenarios}

\begin{figure}[ht]
\centering
\includegraphics[width=0.9\textwidth]{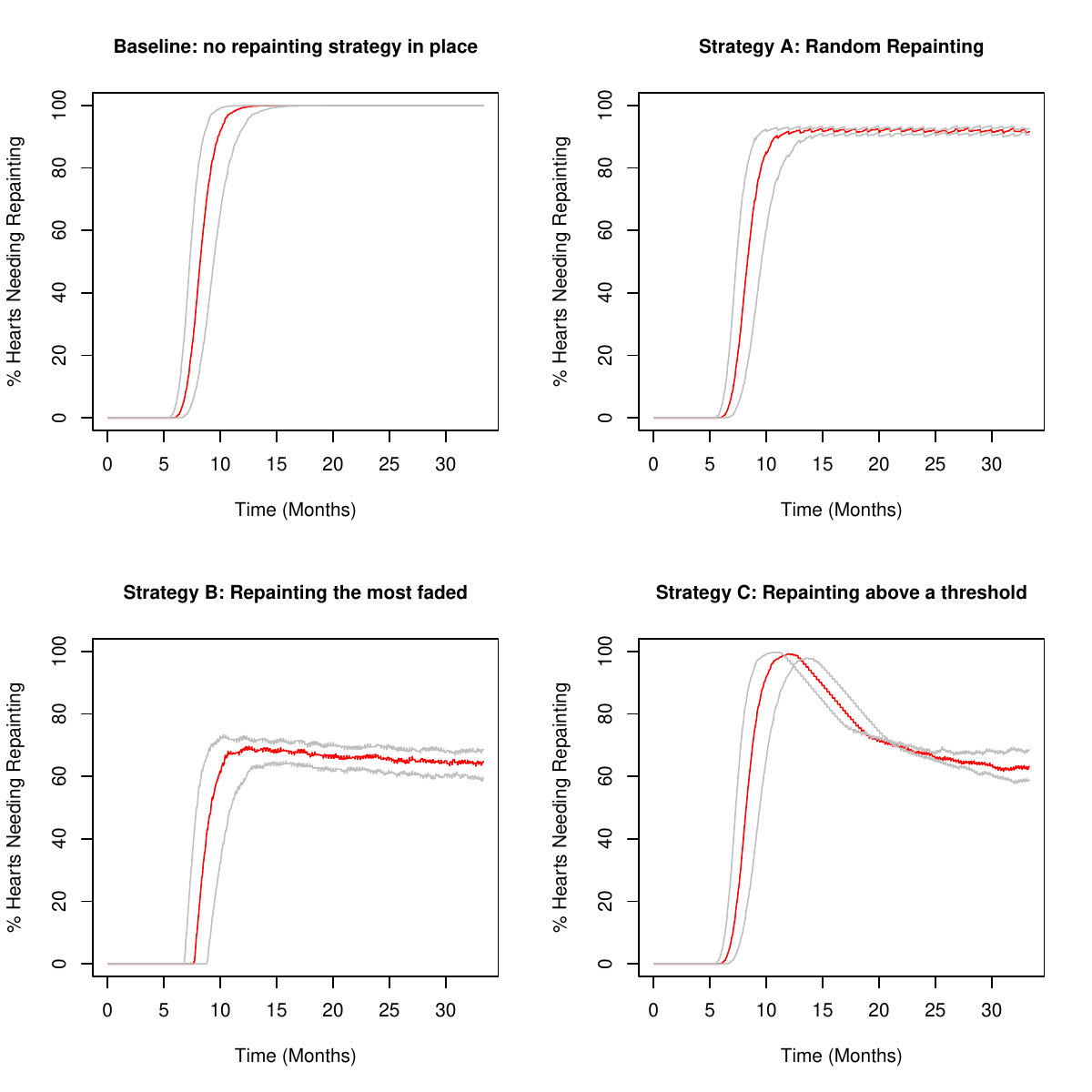}
\caption{Evolution of the collection of hearts assuming a rate of repainting of 5\% of a week (approximately 1000 hearts). The gray lines indicate the 95\% confidence interval.  }\label{fig8}
\end{figure}

Four repainting strategies for the management of the National Covid Memorial Wall were defined utilising Posca markers (Paint I) and considering varying repainting percentages and timeframes. These strategies include no intervention (baseline), random repainting (Strategy A), repainting the most faded hearts (Strategy B), and repainting those that have reached a visible fading threshold (Strategy C). Figure \ref{fig8} shows the \% of hearts that need repainting in each scenario.  The primary purpose of this analysis is not to recommend a specific strategy, but rather to illustrate how this type of modelling facilitates the comparison of multiple strategies. The modelled scenarios are listed below.

\begin{itemize}
    \item \textbf{Baseline:}	No Repainting Strategy in Place. In this strategy, no repainting measures are implemented, allowing the hearts to fade naturally without intervention. Under this baseline condition, all hearts are expected to reach a $\Delta E$ value exceeding 10 within a 15-month period.
    	\item \textbf{Strategy A:} Random Repainting. This strategy involves repainting a fixed percentage of hearts, such as 1\%, weekly. The selection process is random, disregarding the current fading state, meaning that both faded and non-faded hearts may be repainted.  
	\item \textbf{Strategy B:} Repainting the Most Faded. This strategy focuses on repainting the hearts exhibiting the highest degree of fading. A systematic assessment is conducted, either through precise measurement or visual evaluation, to identify and prioritise the most faded 1\% of hearts for repainting. Although this approach may be more effective in maintaining some hearts below the $\Delta E$ threshold of 10, it is not entirely efficient, as numerous hearts may be on the verge of crossing the threshold, and it does not halt the overall fading process.
	\item \textbf{Strategy C:} Repainting above the Fading Threshold. In this strategy, hearts are repainted once they have reached a visibly faded threshold, but the selection process remains random. Instead of relying on a ranking of the hearts in more need, visibly faded hearts are randomly chosen for repainting. Although this method is less precise than Strategy B, it is more practical to implement and can yield similar long-term results. Over time, as the situation is brought under control, the outcomes of Strategy C tend to align closely with those of Strategy B, but with less operational complexity.
\end{itemize}

Numerous alternative strategies could be devised, such as repainting different sections of the wall on a weekly basis, adjusting the repainting percentage according to the extent of fading, or scaling resource allocation based on the level of deterioration. While these alternatives could be explored and outlined in detail within the research, the core message is that such modelling enables collection managers to evaluate and compare the potential impacts of various strategies.
	
    \subsection{The effects of the repainting strategy}
    
\begin{figure}[ht]
\centering
\includegraphics[width=0.9\textwidth]{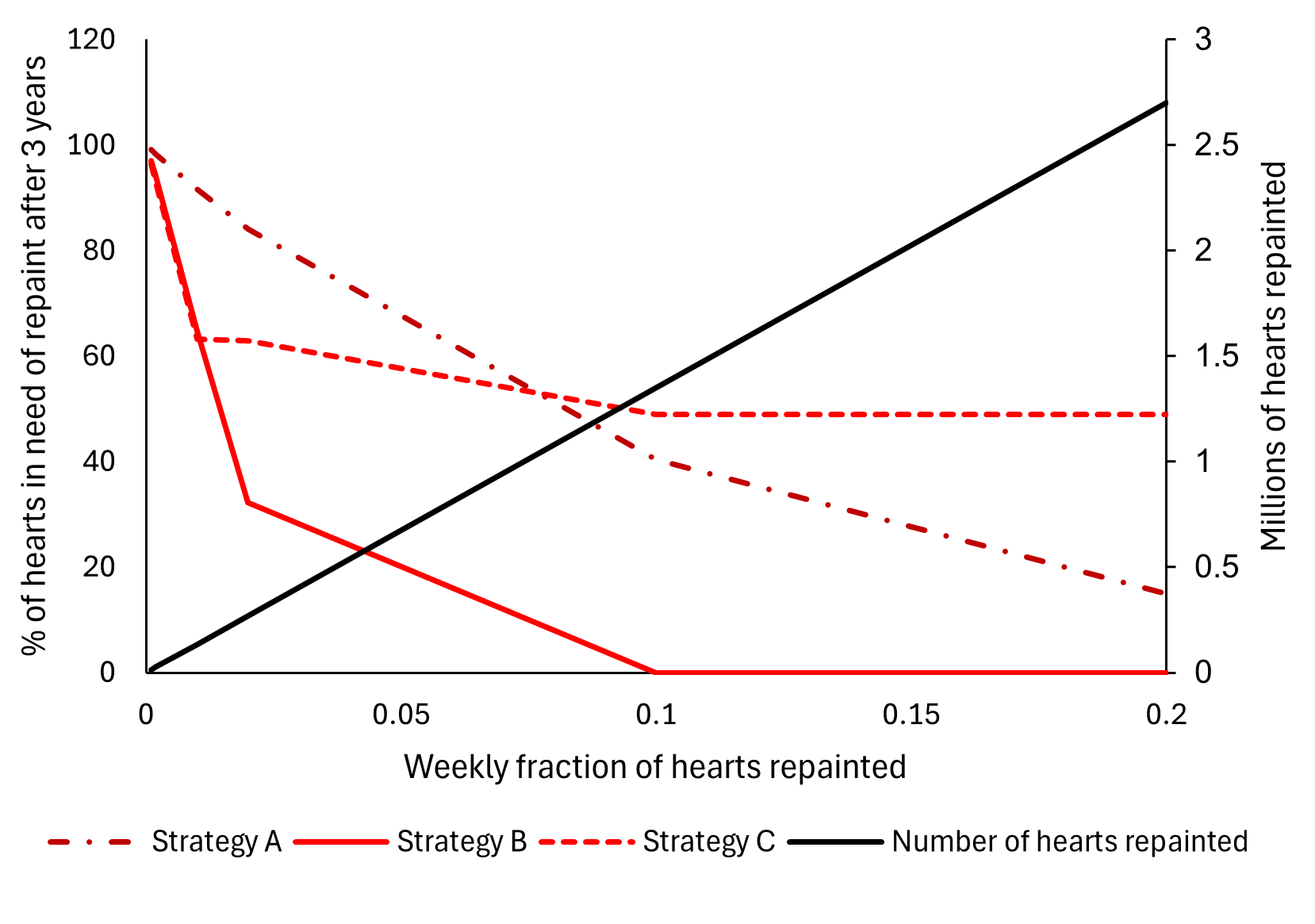}
\caption{Graph to allow decision-making among the repainting strategies. The horizontal axis shows the choice in the fraction of hearts repainted weekly. The vertical axes show the consequences of this decision: how many hearts will be in need of repainting in 3 years, and how many will be repainted in total.  }\label{fig9}
\end{figure}

To illustrate the usefulness of a collections demography, Figure \ref{fig9} was generated to map the controllable variables within this analysis. The primary decision not only involves selecting a strategy but also determining the weekly fraction of hearts that should be repainted.

For example, should 5\% of the hearts be repainted each week? Or would it be more appropriate to repaint 10\%, or even 20\%? Figure \ref{fig9} shows that, depending on the selected fraction, the number of hearts in poor condition or requiring repainting after a period of three years, an arbitrary timeframe set by the researchers, varies. If a 0\% repainting rate is chosen, all the hearts will require intervention in all scenarios. However, with a repainting rate of 0.05\%, the results depend on the strategy. The figure indicates that if the capacity of the volunteers is to repaint 5\% every week, Strategy B is the most effective, followed by C, and finally by A. Strategy A, which is random, only becomes useful when large proportions of hearts are painted.  
 
This analysis provides a method for comparing different strategies. Another interesting result from the graph is that, regardless of the chosen strategy, the same number of hearts are repainted in each scenario. Thus, while the percentage of repainting remains constant, some strategies are more efficient than others, leading to significantly better outcomes when the same resources in terms of paint, personnel, and time are invested.

	\subsection{The benefits of the change in paint}
    
\begin{figure}[ht]
\centering
\includegraphics[width=0.9\textwidth]{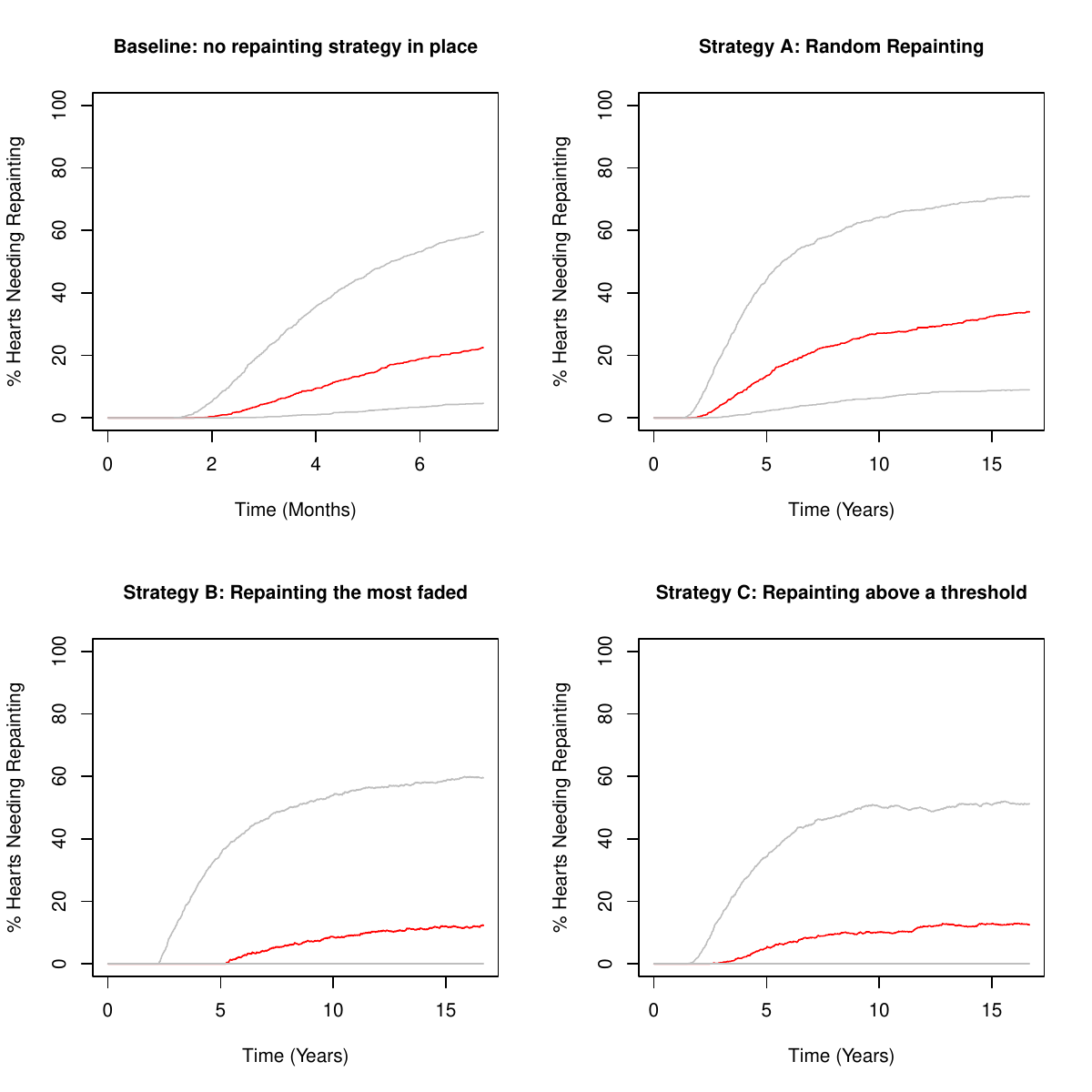}
\caption{Evolution of the collection of hearts assuming a rate of repainting of 1\% of a week (approximately 250 hearts), and assuming a rate of fading of Paint II of 0.5 $\Delta E$/year. The gray lines indicate the 95\% confidence interval.}\label{fig10}
\end{figure}

The used methodology is not precise enough to measure the fading rate of Paint II. However, it is possible to estimate an approximate rate. We know that the manufacturer, Valspar, offers a guarantee that the high quality outdoor paint will not fade or crack in 15 years \cite{Valspar2025}. Considering a conservative threshold of visible fading of 5 $\Delta E$, this would imply a rate of 0.34 $\Delta E$ per year . This corresponds to the rates measured in the fading of high quality architectural finishes exposed for 9 years outdoors in Florida, which ranged from 0.17 to 0.75 $\Delta E$ per year \cite{Wood2018}. Let us assume, then, that the fading rate of the new paint is between these values, 0.5 $\Delta E$/year, with the same uncertainty measured in the case of Paint I (12\% error). In this case, the progression of fading in the three repainting scenarios would follow the curves shown in Figure \ref{fig10}. Not only the collection will last for longer, but a low rate of repainting of 1\% per week will result in the fading stabilising at around 10-20\% of the collection in good condition.  However, due to a longer simulation period, these predictions are subject to a large uncertainty.

\section{Conclusions}

The findings of this study have direct implications for heritage policy, particularly in the context of ongoing public campaigns to secure the National Covid Memorial Wall’s permanent status. The agent-based modelling demonstrates that even with improved materials, long-term sustainability requires structured maintenance and resource allocation. These findings can inform future decisions around official designation. 

The decision to transition from Posca paint markers (Paint I) to Valspar Premium Masonry paint (Paint II) was a prudent one. The rapid fading of the Posca paint required an unsustainable rate of repainting, which placed a significant burden on the volunteers. The simulation based on the rough estimation of the fading rate for Paint II, suggests that repainting efforts will still need to be maintained at approximately 250 hearts per week. This is essential to keep a stable and low percentage of hearts perceived as needing repainting.

This finding has important implications for the long-term preservation and memorialisation of the National Covid Memorial Wall. Despite the improved durability of Paint II, the ongoing need for substantial repainting suggests that any long-term, sustainable management strategy must involve either a provision for regular maintenance or a further reduction of the fading rate. 

This study relies on publicly shared images and volunteer engagement to inform the conservation of a highly emotive heritage site. While the use of citizen science offers significant methodological and participatory benefits, it also raises ethical considerations. In particular, the emotional proximity of volunteers and social media users to the memorial demands a sensitive approach. A citizen science framework does not preclude that consent and agency are prioritised, like any research with participants. Any future research must continue to respect the memorial’s role as a space of mourning and political expression.

From a methodological perspective, this study demonstrates the feasibility of using a collections demography approach, supported by citizen science and social media data, to inform heritage management decisions. The methodology allowed for the collection of extensive data on the fading rates of the hearts, providing a sufficiently solid basis for modelling decision-making scenarios. However, the reliance on publicly available images and volunteer-collected data introduces variability and potential biases that must be considered. Naturally, these estimations would be better with more controlled data collection protocols and the use of more precise measurements. Without altering the citizen science principles applied in this research, it would be possible to obtain better data by:

\begin{itemize}
    \item Inviting submissions from the public, rather than relying on uncontrolled social media images
	\item Including a sign on site that acts as a prompt to participate as well as an image calibration chart
	\item Doubling on the efforts with the volunteers who, acting as citizen scientists, can provide longitudinal data on fading
\end{itemize}
	
We hope this study illustrates the flexibility of the collections demography framework: it can be used not only with high-quality, research-grade data, but also, as we show here, with lower quality data from social media and citizen science.

\section*{Acknowledgements}

Thank you to the students of the MSc Sustainable Heritage who contributed to this work over several years: Cheng Alex, Chang Yizhang, Chen Yining, Du Yiming, Fan Yuqing, Gao Yunfei, Guo Diya, Hu Zhijing, Huang Yuching, Li Chengxi, Luo Jia, Miao Runjie, Shen Ke, Sun Enling, Wan Haocheng, Yin Yihan, Zhang Yi. The authors are also thankful to the Friends of the Wall, for their interest and warm welcome. 

\bibliography{sn-bibliography}

\end{document}